\documentclass{article}
\usepackage{microtype}
\usepackage{graphicx}
\usepackage{subfigure}
\usepackage{booktabs} 
\usepackage{hyperref}
\usepackage{multirow}
\usepackage{booktabs}
\usepackage{multirow}
\usepackage{booktabs}
\usepackage{adjustbox}
\usepackage{array}
\usepackage{makecell}
\usepackage{graphicx}
\usepackage{xcolor}

 \usepackage[accepted]{mlsys2025}

\mlsystitlerunning{C-FedRAG:  Confidential Federated RAG}

\begin{document}

\twocolumn[
\mlsystitle{C-FedRAG: A Confidential Federated Retrieval-Augmented Generation System}


\mlsyssetsymbol{equal}{*}

\begin{mlsysauthorlist}
\mlsysauthor{Parker Addison}{dlt} 
\mlsysauthor{Minh-Tuan H Nguyen}{dlt} 
\mlsysauthor{Tomislav Medan}{dlt} 
\mlsysauthor{Jinali Shah}{dlt}
\mlsysauthor{Mohammad T Manzari}{dlt} 
\mlsysauthor{Brendan McElrone}{dlt} 
\mlsysauthor{Laksh Lalwani}{dlt} 
\mlsysauthor{Aboli More}{dlt} 
\mlsysauthor{Smita Sharma}{dlt} 
\mlsysauthor{Holger R. Roth}{nvda} 
\mlsysauthor{Isaac Yang}{nvda} 
\mlsysauthor{Chester Chen}{nvda}  
\mlsysauthor{Daguang Xu}{nvda}  
\mlsysauthor{Yan Cheng}{nvda}  
\mlsysauthor{Andrew Feng}{nvda}  
\mlsysauthor{Ziyue Xu}{nvda}  
\end{mlsysauthorlist}

\mlsysaffiliation{dlt}{Deloitte, New York, NY, USA}
\mlsysaffiliation{nvda}{NVIDIA, Santa Clara, CA, USA}

\mlsyscorrespondingauthor{Ziyue Xu}{ziyuex@nvidia.com}
\mlsyscorrespondingauthor{Mohammad Manzari}{mmanzari@deloitte.com}

\mlsyskeywords{Federated Learning, Retrieval Augmented Generation, Confidential Computing}

\vskip 0.3in

\begin{abstract}
Organizations seeking to utilize Large Language Models (LLMs) for knowledge querying and analysis often encounter challenges in maintaining an LLM fine-tuned on targeted, up-to-date information that keeps answers relevant and grounded. Retrieval Augmented Generation (RAG) has quickly become a feasible solution for organizations looking to overcome the challenges of maintaining proprietary models and to help reduce LLM hallucinations in their query responses. However, RAG comes with its own issues regarding scaling data pipelines across tiered-access and disparate data sources. In many scenarios, it is necessary to query beyond a single data silo to provide richer and more relevant context for an LLM. Analyzing data sources within and across organizational trust boundaries is often limited by complex data-sharing policies that prohibit centralized data storage, therefore, inhibit the fast and effective setup and scaling of RAG solutions. In this paper, we introduce Confidential Computing (CC) techniques as a solution for secure Federated Retrieval Augmented Generation (FedRAG). Our proposed Confidential FedRAG system (C-FedRAG) enables secure connection and scaling of a RAG workflows across a decentralized network of data providers by ensuring context confidentiality. We also demonstrate how to implement a C-FedRAG system using the NVIDIA FLARE SDK and assess its performance using the MedRAG toolkit and MIRAGE benchmarking dataset.

\end{abstract}
]

\printAffiliationsAndNotice{}  

\section{Introduction}
\label{intro}
\subsection{Background and Motivation}
Much of the excitement about LLMs and Generative AI has come from their potential to quickly retrieve and analyze information from both structured and unstructured data across different sources \cite{ghodratnama2023adaptingllmsefficientpersonalized}. Despite their success in many applications as models pre-trained on historical data, major concerns from users include: 1) hallucinations in answers without providing real evidence, and 2) limitations in utilizing relevant information sources that are frequently updated \cite{lewis2021retrievalaugmentedgenerationknowledgeintensivenlp}. RAG has emerged as a popular method that offers significant advantages and improvement over pre-trained models for grounding LLMs in factual information \cite{shuster2021retrievalaugmentationreduceshallucination} . There have been many academic and industry publications pointing to the role and impact of RAG-based systems in reducing LLM hallucinations, improving the accuracy of results with new contextual information, and providing an alternative to computationally expensive fine-tuning \cite{li2024enhancingllmfactualaccuracy, chen2024llmstuningmethodswork}. 
 
Nevertheless, proof-of-concept LLM applications using RAG-based techniques can face challenges, including barriers to data access and ineffective retrieval methods. Moreover, many RAG-based systems are designed with the assumption of a centralized data architecture and relatively unfettered access to the required data by the LLM application \cite{xu2024genaipoweredmultiagentparadigmsmart}. As the key component of a RAG system, the contextual information yielded by the retrieval determines the quality of the final output. Therefore, to unlock the full potential of RAG, users may want to fetch information from many relevant sources, rather than being limited by a siloed dataset. \cite{hardjono2017openalgorithmsidentityfederation, jiangmit2024}. With preliminary works investigating such distributed systems on several applications, one critical concern has not been properly addressed: the collection of contexts from various sources can violate data security rules and essentially break one of the key benefits of such systems. In this paper, we introduce \textbf{C-FedRAG} (Confidential Federated RAG), a secure federated RAG system that integrates CC to enhance the privacy of sensitive contexts. C-FedRAG enables effective scaling of LLMs applications, unlocks otherwise isolated data sources, and enhances the performance of RAG systems without compromising security through decentralized information retrieval and synthesis within a secure environment.

\subsection{Related Work}
To situate our work within the broader context of existing research around Federated Learning (FL) with RAG systems, we highlight some related works:

FeB4RAG: Evaluating Federated Search in the Context of Retrieval Augmented Generation \cite{wang2024evaluatingfederatedsearch} outlines an architecture for federated search within RAG frameworks and introduces a new dataset for evaluating federated search, addressing the limitations of existing collections. The paper emphasizes the importance of developing sophisticated federated search strategies to optimize RAG pipelines and enhance the quality of generated responses.

Federated Learning-Enhanced Retrieval Augmented Generation (FLERAG) \cite{kim2024federatedlearningenhancedrag} proposes a new approach for selecting the best response between a traditional RAG LLM and an FL model trained across client device data, in order to solve the concern of needing to continuously update the RAG database. Selection is done through a response arbiter which selects the response with higher confidence. The global FL model is shared across all clients, providing a more comprehensive and up-to-date response for comparison with the LLM response that was based on pre-trained knowledge.

Cache Me If You Can: The Case For Retrieval Augmentation (RA) in Federated Learning \cite{muhamed2024cachemeifyoucan} proposes an approach that uses RA to enhance FL by incorporating a retrieval-based method during inference, where the client's device retrieves relevant information from its local dataset and augments the query before feeding it to the model. This addresses privacy concerns and regulatory compliance while allowing individual clients' models to benefit from the collective knowledge of the entire network. The approach necessitates that clients in the federated learning system fine-tuned their own models using their respective private data.

Clinical Question-Answering over Distributed EHR
Data \cite{jiangmit2024} proposed the use of federated RAG for Clinical QA, leveraging LLMs for clinical question answering without compromising patient privacy. The proposed system utilized a hierarchical design for federated document retrieval, enabling efficient and secure access to distributed clinical data. The author also introduced a new dataset, based on the MIMIC-IV database, specifically designed for evaluating clinical QA systems. By addressing privacy concerns and enhancing interpretability, the proposed approach represented a significant step forward in leveraging LLMs for clinical applications.

\subsection{Contributions}
Building upon the related works described previously, our proposed  \textbf{C-FedRAG} system architecture introduces computational workflows that execute RAG in a federated fashion, using Confidential Computing (CC) techniques to comply with data privacy and security constraints. For instance, the embedding models used to convert raw text chunks into vectors could be trained across multiple data providers using FL. The context retrieval and re-ranking steps could be implemented in a federated fashion across a network of data providers. The final inference step in RAG, where the original user query and retrieved contexts are combined and used to generate a response, could also involve a federated computation across different foundation model providers. All these computations and information flows can be secured by utilizing CC techniques to ensure reliable data collection and compliance with relevant regulations.

\begin{figure*}[hbt!]
    \centering
    \includegraphics[width=\linewidth]{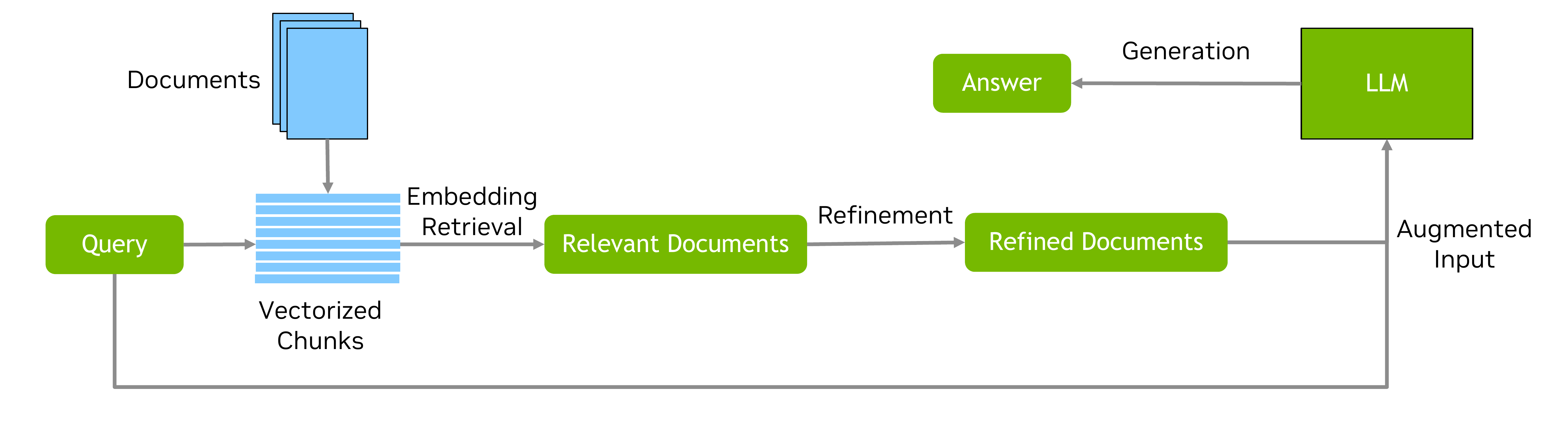}
    \caption{RAG pipeline illustration}
    \label{fig:rag}
\end{figure*}

We believe that C-FedRAG holds promise in helping organizations scale LLM applications more effectively. Take an industry application with a single-entity and multiple divisions for example. It is a common scenario where enterprise data is scattered across multiple business units and IT systems. This heterogeneous data environment makes it difficult to centralize the data for retrieval purposes, let alone to integrate data from different sources and modalities \cite{konečný2016federatedoptimizationdistributedmachine}. C-FedRAG will be able to address these challenges within an enterprise through its distributed nature. Moreover, with secure mechanisms, C-FedRAG can help organizations collaborate with each other by enabling trusted data flows between partners. 

The key contributions of this work are:
\begin{itemize}
    \item The design and implementation of a fully federated system for performing RAG across decentralized data providers
    \item A novel orchestrator architecture using Confidential Computing Environments to safe-guard privacy and security of contextual data.
    \item A comprehensive evaluation of the confidential federated RAG system's ability to handle various medical text formats and complexities.
\end{itemize}

\section{Method}
In this section, we first start with the basic flow of RAG, then we introduce concepts of C-FedRAG by visiting the key stages of a RAG pipeline and potential ways to improve each step with federated computation, and conclude with implementation details.

\subsection{Flow of RAG}
As shown in Figure~\ref{fig:rag}, the basic RAG flow can be described as the following: 
\begin{enumerate}

\item	To prepare for the retrieval, an offline process is first performed over a given database to formulate the documents into vectorized chunks where each chunk is represented by an embedding vector. The step is done by an ``embedding model''. 
\item	When a query is entered, it is first embedded by the same embedding model as the respective data, and embedding retrieval is performed such that the top-k most relevant document chunks will be retrieved according to the query. These top-k relevant documents are usually retrieved based on a distance metric in the embedding space. 
\item	Since the embedding retrieval step is highly dependent on the performance and relevance of the embedding model, as well as the distance metrics being used, the results may not be optimal. Therefore, a refinement step is often necessary to further polish the retrieval result from the last step. One example of such refinement is to use a re-ranking model to estimate the relevance score w.r.t. the query for each of the Top-k relevant documents. As the ranking model is trained to directly estimate the relevance of a pair of inputs, rather than relying on their embedding distance, it can be more accurate for many applications. 
\item	Once the top-k relevant documents are further refined into the top-n candidates, they will then be used as context and appended to the original query to form the augmented input. This augmented version of the query will be fed to the LLM for generating the response. Since it contains richer information than the plain query, the answer can be more accurate and relevant to the given context \cite{wang2024bioragragllmframeworkbiological}. 

\end{enumerate}

Given the RAG system flow in Figure~\ref{fig:rag}, we can notice that the process has 3 key pieces: 
\begin{itemize}
    \item The embedding model to vectorize a given database, and to provide the rough retrieval results
    \item The refinement / re-ranking model to polish the retrieved documents
    \item The LLM model to generate answers to the augmented query
\end{itemize}

All 3 pieces would have special considerations under a federated setting and security specifications. 

\begin{figure*}
    \centering
    \includegraphics[width=.8\linewidth, height=.3\textheight,keepaspectratio]{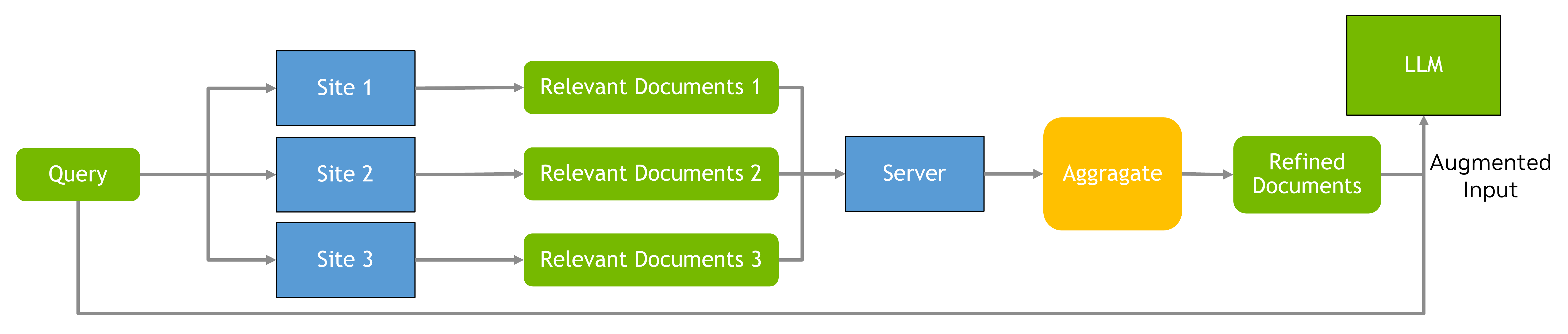}
    \caption{Confidential Federated RAG pipeline}
    \label{fig:fedrag}
\end{figure*}

\subsection{Key Concepts of Confidential Federated RAG}
In a federated RAG system, components include a user interested in finding the answer to a query, several data providers offering relevant documents, and an orchestrator for routing, fusing, and protecting the information flow for all parties. As shown in Figure~\ref{fig:fedrag}, the basic flow of a C-FedRAG system is as follows:
\begin{enumerate}
    \item Similar to local RAG, each data provider will first vectorize their dataset using an embedding model. Note that this embedding model can be the same across all parties, or customized to each provider's own data distribution.
    \item When a user initiates a query, the orchestrator will evaluate the query and forward it to all or a selected subgroup of the data providers. 
    \item Every data provider who receives the query will perform a local retrieval process. Similar to local RAG, this could involve an embedding retrieval plus a refinement step. The resulting (refined) relevant documents will be sent back to the orchestrator,in the form of raw text, or embeddings depending on the security requirements and later system design. 
    \item Upon receiving the responses from data providers, the orchestrator will perform an aggregation step in a CC environment to combine (and refine) all contexts from different providers (e.g. ``re-ranking''), and the aggregated context will be used to produce the augmented input.
    \item Once the orchestrator generates the augmented input with the aggregated context from several data providers, it will forward it to the LLM for inference in a CC environment, and will send the received answer back to the user. 
\end{enumerate}

Most importantly, as the orchestrator receives raw contexts from all providers, this poses significant concerns over data leakage. Therefore, we would want to reinforce secure mechanisms, like CC Environments, to ensure data security during the three orchestration operations: context collection, context aggregation, and response generation.  
The above flow can be summarized as Algorithm~\ref{alg:fedrag}. Under a federated setting, we can have several variations for embedding/ranking models, context retrieval and aggregation, and the overall retrieval-generation process. 
 
For the embedding and re-ranking models, the most straightforward solution is to use a single model for all parties with an arbitrary off-the-shelf model. However, these pre-trained models may not fit the local databases well and can produce sub-optimal results. Therefore, it can be preferable that specialized embedding and ranking models be trained over the data-providers’ data distribution. Common strategies of FL can be adopted in training these models, including but not limited to: basic FL where a global model will be trained over each participant's data, and personalized FL, where each participant will have a personalized model at the end of training, fitting to its own data distribution while taking advantage of the global common information. 
 
For context retrieval and aggregation, the simplest form is to use standard local retrieval based on embedding and ranking models, and combine the contexts from all data providers by simple concatenation. However, if the number of data providers is large, simple concatenation of the responses can quickly drain the token budget for generating the augmented input to an inference LLM and reduce output accuracy \cite{pmlr-v30-Wang13}. Thus, more sophisticated refinement methods will be of critical importance to reduce the information redundancy at the aggregation stage. One basic solution is to use a ranking model to perform a round of global ``re-ranking'', such that contexts received from all providers are ranked again, and the most relevant ones will be selected for the final generation. Again, due to the data-sensitive nature of this step, collecting raw context, secure mechanisms like CC are needed to ensure data security and privacy.

\begin{algorithm}[tb]
   \caption{C-FedRAG Algorithm}
   \label{alg:fedrag}
\begin{algorithmic}
   \STATE {\bfseries Input:} user query $x$, local context size $m$, global context size $n$, site number $k$
   \STATE {\bfseries Models:} embedding model $F_{emb}$, aggregation model $F_{aggr}$, inference model $F_{inf}$
   \STATE 
   \STATE Initialization: vectorize local corpora to $V_i$ with $F_{em}$.
   \FOR{each $x$}
   \STATE Distribute $x$ to a number of sites $k_n \leq k$
   \FOR{$i=1$ {\bfseries to} $k_n$}
   \STATE Site-$i$ retrieves $m$ relevant contexts with distance metrics $Dist(F_{em}(x), V_i)$ 
   \STATE Return $m$ contexts $C_{local_i}$ to orchestrator
   \ENDFOR
   \IF{Within secure (CC) environment}
   \STATE Local contexts to global context by aggregation model $C_{global} = F_{aggr}(C_{local_1}, ..., C_{local_{k_n}})$
   \STATE Generate response $y = F_{inf}(x, C_{global})$
   \ENDIF
   \ENDFOR
\end{algorithmic}
\end{algorithm}

For the overall retrieval-generation process, a basic flow includes broadcasting the query to all parties and running inference with a single LLM for final answer generation. Beyond such a baseline flow, advanced techniques can help the federated flow for better RAG performance. When distributing the query, instead of blindly broadcasting to everyone, a selective process can be added to only query the most relevant data providers according to the global knowledge of query-provider compatibility. Likewise, before sending the query to a data provider, the query can be pre-processed (re-writing, expansion, etc.) in a personalized fashion. On the other end of the process, instead of one single global LLM for generation, we can employ multiple model providers, who can be the same parties as data providers, or not. Similarly, a routing model can orchestrate the answer inference by sending the augmented query to the most relevant LLMs, and produce the final answer by aggregating the responses from them.  
 
Considering the above potential variations, we have chosen a basic C-FedRAG system setup as follows:

\begin{enumerate}
    \item The orchestrator (within CC environment) broadcasts a user query to all data providers. 
    \item Data providers perform standard local retrieval by using a pre-trained, off-the-shelf embedding model. 
    \item Upon receiving the locally retrieved relevant documents, the orchestrator securely aggregates them (within CC environment) by performing a re-ranking with a pre-trained ranking model. 
    \item The aggregated context is concatenated with the query in a prompt template for answer generation via a single LLM running in CC environment.  
\end{enumerate}

\subsection{Implementation}
\begin{figure*}
    \centering
    \includegraphics[width=\linewidth]{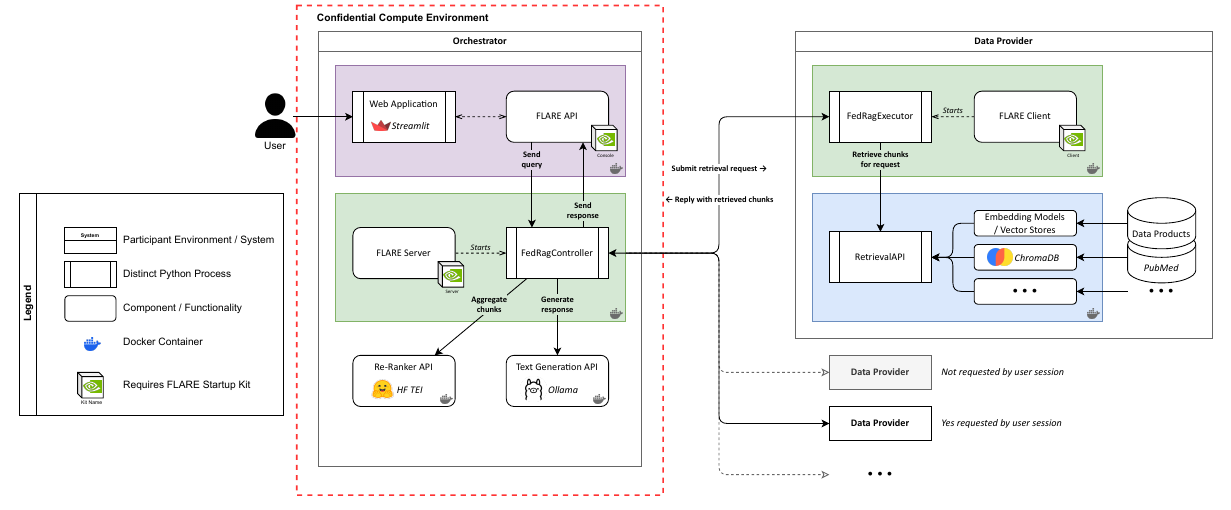}
    \caption{C-FedRAG implementation}
    \label{fig:impl}
\end{figure*}

To implement the above C-FedRAG, we embraced a federated-native approach by building the solution around NVIDIA FLARE (NVFlare) \cite{https://doi.org/10.48550/arxiv.2210.13291}, a federated application framework. The implementation details are illustrated in Figure~\ref{fig:impl}.

\subsubsection{Secure Communication}
In the C-FedRAG system, secure communication is handled by NVFlare where a strict client-server relationship is enforced, including specific port requirements for both the FL Server and FL Clients. This secure, client-server architecture provides several benefits:
\begin{itemize}
    \item \textbf{Centralized Management:} All requests are directed to the server, allowing the server to enforce security policies, access controls, and rules across all clients.
    \item \textbf{Security:} Mutual Transport Layer Security (mTLS) protocol is required for all connections between the server and clients.  This protocol not only ensures all communication is encrypted, but also requires two-way identity verification.  That is, the server has to present its X.509 certificate to clients for authentication and clients have to present their X.590 certificates to the server for authentication.  The authentication process is performed by establishing the entire X.509 certificate chain on the server and clients.
    \item \textbf{Data Privacy:} Filters are used on pre-process and post-process tasks for data privacy. These filters are applied to both task data and task results on both the server and client sides. Filters are powerful tools that allow the system to control and protect data transmission.
\end{itemize}

The client-server architecture is particularly effective for applications that require centralized control, security, and data privacy. The C-FedRAG system, with this strict communication strategy, is designed so that each participant in the system has maximal flexibility and privacy.

\subsubsection{Orchestrator}
The orchestrator uses an NVFlare job to coordinate retrieval requests from each of the selected data providers and collects all retrieved chunks into a list. The list of retrieved chunks is then passed to a re-ranker model for down-selection, selecting only the n-most relevant chunks. The re-ranker model, \textit{bgre-reranker-base}, is a transformer-based model that directly takes in two sequences and outputs a score of how ``relevant'' the latter is to the prior \cite{bge_embedding}. Scores are computed pairwise across the query and all candidate retrieved chunks, then the chunks with the highest relevance scores are used as the final chunks. Upon completion of down-selection, the original query and final relevant chunks are passed as a prompt to a text generation model producing the final answer. For final inference, we used Llama3-8B-Instruct as the LLM and the text generation model. This is all done within the orchestrator wrapped inside a CC environment.

\subsubsection{Confidential Computing}
NVFlare facilitates the secure retrieval and synthesis of information during communication. However, to ensure and protect the confidentiality and integrity of raw data at the orchestrator, we integrated a privacy-preserving approach for retrieval/aggregation and LLM inference within the C-FedRAG workflow using confidential virtual machines \cite{ConfidentailComputingConsortium}, which are fully isolated virtual machines with additional protection like cryptographic attestation for securely processing highly sensitive data and ensuring that only authorized codes are running. The architecture of the C-FedRAG solution leverages such CC environment to ensure the privacy of contextual data processed at the orchestrator by RAG workflows. By incorporating CC with a Trusted Execution Enclave integration, data providers' contributed data remains confidential in the enclaves during both retrieval and inference processes \cite{lee2024privacypreservingdecentralizedaiconfidential}. This implementation not only fortifies the security of the data but also upholds the integrity and privacy standards required for sensitive information handling in FL systems.

\subsubsection{Data Providers}
To perform the federated retrieval, each data provider is equipped with a standardized API interface to perform embedding and retrieval from a VectorDB using LlamaIndex. The data providers are required to implement methods for listing available data products and retrieving chunks from requested products when provided with a query. In our implementation, we use META AI Research's Contriever \cite{https://doi.org/10.48550/arXiv.2112.09118} with a Chroma vector database. The Contriever embeddings model is designed to learn vector representations of text data, where semantically similar chunks of text are encoded as points that are close together in the embedding space. This is achieved by using a contrastive loss function that encourages similar pairs to be close and dissimilar pairs to be far apart. Essentially, the query gets embedded by each data provider using its embedding model of choice, and the top-k chunks with the closest distance of their pre-computed embeddings are returned.

\section{Experiment Results}
The MedRAG toolkit was used for our experimentation design and the MIRAGE dataset was the foundation of our benchmarking approach \cite{xiong2024benchmarkingretrievalaugmentedgenerationmedicine}. More specifically, we ran experiments using a fixed set of corpora, embedding/retrieval algorithms, context windows, models, and QA datasets.

\begin{table*}
\centering
\begin{tabular}{llccc}
\hline
LLM & Method & PubMedQA & BioASQ-Y/N & Avg. \\
\hline
\multirow{3}{*}{\makecell[l]{LLaMA-3- \\ 8B-Instruct}} 
 & CoT & 54.03 $\pm$ 2.23 & 65.52 $\pm$ 1.91 & 59.78 \\
 & MedRag (MedCorp) & 67.20 $\pm$ 2.10 & 74.60 $\pm$ 1.75 & 70.90 \\
 & MedRag (PubMed) & 67.20 $\pm$ 2.10 & 74.27 $\pm$ 1.76 & 70.74\\
  & MedRag (StatPearls) & 41.60 $\pm$ 2.20 & 55.34 $\pm$ 2.00 & 48.47 \\
   & MedRag (Textbooks) & 36.40 $\pm$ 2.15 & 55.66 $\pm$ 2.00 & 46.03 \\
    & MedRag (Wikipedia) & 42.60 $\pm$ 2.21 & 67.15 $\pm$ 1.89 & 54.88 \\
 & C-FedRAG (Embedding Rank) & 61.60 $\pm$ 2.18 & \colorbox{green!25}{76.05 $\pm$ 1.72} & 68.83\\
  & C-FedRAG (Re-rank Model) & 65.40 $\pm$ 2.13 & \colorbox{green!25}{79.61 $\pm$ 1.62} & \colorbox{green!25}{72.51} \\
\hline
\end{tabular}%
\caption{\centering Benchmark results comparing C-FedRAG to vanilla RAG results (MedRAG) and non-RAG LLM prompting}
\label{tab:benchmark-results}
\end{table*}
\vspace{0pt}

\begin{table*}
\centering
\begin{tabular}{llccc}
\hline
LLM  & PubMedQA & BioASQ-Y/N & Avg. \\
\hline
LLaMA-3-8B-Instruct & 54.03 $\pm$ 2.23 & 65.52 $\pm$ 1.91 & 59.78 \\
LLaMA-3.1-8B-Instruct & 57.60 $\pm$ 2.21 & 78.64 $\pm$ 1.65 & 68.12 \\
LLaMA-3.2-1B-Instruct & 48.60 $\pm$ 2.24 & 57.12 $\pm$ 1.99 & 52.86 \\
LLaMA-3.2-3B-Instruct & 51.00 $\pm$ 2.24 & 72.49 $\pm$ 1.80 & 61.75 \\
\hline
\end{tabular}%
\caption{\centering Ablation of inference LLM with latest LLaMA models using CoT}
\label{tab:benchmark-llm}
\end{table*}
\vspace{0pt}

\subsection{Corpora}
For test contextual documents, we collectively used a subset of four datasets as context providers: Pubmed, Wikipedia, Statpearls and 
Textbooks, which are available from the MedRAG toolkit. The top 10,000 scoring snippets in each corpus for each query task were pulled and embedded to create our corpora subsets. These four datasets were distributed between two simulated data providers, where Pubmed and Wikipedia were housed in Site 1 and Statpearls and Textbooks were housed in Site 2. 

\subsection{Embedding and Retrieval}
We utilized the Contriever embedding and retrieval algorithm from Facebook Research \cite{https://doi.org/10.48550/arXiv.2112.09118} to embed each of the aforementioned datasets. Each of the embedded datasets was stored using a FAISS vector index after a one-time embedding procedure. Retrieval was conducted via the Contriever algorithm at each of the mock sites (Site 1, Site 2) where the top 8 chunks from each dataset was retrieved and pulled back to the orchestrator. 

\subsection{Context Window}
Each final LLM input had a context window of 8 final chunks. After the aforementioned 8 chunks were retrieved from each dataset, reranking was done on the 32 total chunks via the BAAI/bge-reranker-base \cite{bge_embedding} re-ranker algorithm to produce a final 8 chunks to be used in the context window. 

\subsection{Model}
Using Ollama, we ran the Llama3-8B-Instruct 4 bit quantized model to generate our LLM responses after the re-ranked chunks were appended to the system prompt to create each context window for each query.

\subsection{QA Datasets}
We kept the scope of our queries to the BioASQ and PubMedQA datasets. BioASQ contains 618 Y/N questions and PubMedQA contains 500 yes/no/maybe questions. 

\subsection{Evaluation}
The totality of the responses to the BioASQ and PubMedQA tasks were passed through the MIRAGE evaluation platform to generate a percentage accuracy and standard deviation score against the ground truth answers. In our evaluation, we covered following scenarios:
\begin{enumerate}
    \item CoT (Chain of Thought): Prompting a baseline LLM without RAG to think through its solution to the question 
    \item MedRag (Medcorp): Vanilla RAG via the MedRag system using a singular dataset combining all four datasets.
    \item MedRag (PubMed): Vanilla RAG via the MedRag system using only the PubMed corpus.
    \item MedRag (StatPearls): Vanilla RAG via the MedRag system using only the StatPearls corpus.
    \item MedRag (Textbooks): Vanilla RAG via the MedRag system using only the Textbooks corpus.
    \item MedRag (Wikipedia): Vanilla RAG via the MedRag system using only the Wikipedia corpus.
    \item C-FedRAG (Embedding Rank): RAG via the C-FedRAG system where chunks from all four data sources were ranked for context window use only through embedding/retrieval rankings through Contriever.
    \item C-FedRAG (Re-rank Model): RAG via the C-FedRAG system where chunks from all four data sources were ranked for context window use through embedding/retrieval rankings and a final reranking via bge-reranker-base.
\end{enumerate}

\subsection{Results}
As shown in Table~\ref{tab:benchmark-results}, context from individual datasets may not be able to provide relevant information as compared with centralized. For the candidate task and dataset, the PubMed dataset dominates the vanilla RAG system, while other datasets have little chance to bring novel contexts - only slightly for the task of BioASQ. C-FedRAG, on the other hand, is able to gather useful information from all sites, and with the help from re-ranking model, it is able to help promoting the performance of BioASQ significantly.

We further performed ablation over the inference LLM using CoT, the results are shown in Table~\ref{tab:benchmark-llm}. As illustrated, RAG can significantly boost the generation accuracy of base model to match and beat that of more recent and state-of-the-art models.

\section{Discussion}
\subsection{Assumptions and Limitations}
The implementation of C-FedRAG is guided by several high-level principles designed to ensure data privacy, control, and security. Our approach sends user's queries to an LLM that is extended and optimized by RAG. Retrieval is performed at each isolated Data Provider ensuring that retrieval is based solely on the provider respective data and allowing the LLM, through an orchestrator, to reference ground truth information from a private knowledge base. The orchestrator coordinates retrieval across data providers, aggregates the final set of retrieved contexts, and generates a final answer using an LLM. The models used for retrieval, re-ranking, and text generation are entirely configurable and customizable based on each data provider's policies. This setup allows data providers to choose their own implementation of retrieval with additional approaches and privacy layers that best fit their data and governance and avoids the need to centralize data sources across multiple data providers. Furthermore, it ensures secure communication as the data providers never communicate with each other and never receive inbound communication. All communication is directed into and routed through the orchestrator running in a CC environment, which never initiates contact.

In this paper, we demonstrate that a secure federated workflow using the NVFlare SDK with CC, effectively allows the scaling of LLM applications within and across enterprises without centralizing data while ensuring secure collaboration. NVFlare ensures communication security via its protection features, and the protection for the raw data from the orchestrator is implemented with CC. CC is essential to prevent sensitive data from being misused by the orchestrator. Enclosing the full orchestrator processes within a CC environment ensures that intermediate results  remain inaccessible.

As a proof-of-concept study, we limited our focus to the text only, RAG specific task, leaving other topics for future studies (multi-modal, other tasks with multi-agent). Instead of having multiple LLMs from different sites, we limit only having an LLM on the aggregator site, while all other sites only retrieve text responses based on the context. 

\subsection{Identity and Access Management} 
 
Federated RAG is still a relatively new concept, and there are still many questions that must be answered before Federated RAG-based systems, including C-FedRAG, can be deployed successfully in enterprise environments. For example, how can user identity and access be seamlessly managed across multiple data providers and security domains? How will access rules for specific data products be defined and broadcast across the network of data providers? Integrating federated identity and access management is a non-trivial task and requires a standardized process for provisioning, harmonization, and brokering identities. 

\subsection{Privacy and Security} 
C-FedRAG systems should also be subject to comprehensive cyber and privacy threat modeling. For example, it is important to understand how we can prevent data poisoning attacks during the training of a federated embedding model. Similarly, we may want to better understand what privacy-enhancing technologies (e.g. differential privacy, trusted execution environments) can be used to shield retrieved context from an ``over-curious'' orchestrator during the LLM inference step. 
 
In addition, if the data providers participating in a C-FedRAG system choose to encrypt retrieved context prior to sending it to the orchestrator, we may need to better understand how to securely store, manage, and exchange multiple encryption keys across the network of data providers. 

\subsection{Context Aggregation and Inference} 
 
One of the most important and exciting research directions for C-FedRAG is related to the aggregation of context retrieved from multiple data providers. In the context of C-FedRAG, it can be seen as similar to the re-ranking process used in other RAG-based workflows. 
In a federated computation process, it may also be desirable to weigh contributions based on the data provider’s credibility or reputation. Aggregation algorithms could be as simple as concatenation or may involve more complex sorting and even summarization operations. 

For inference, the generation LLM can be both global (access by server/orchestrator), or locally hosted by model providers - acting as an ``internet of agents''.

\subsection{Key Benefits}
With the C-FedRAG design, we would like to elaborate further on its key benefits.

First, C-FedRAG can leverage specialized knowledge that can be restricted otherwise. With the distributed nature of many proprietary datasets that closely related to a user's query, at retrieval stage, C-FedRAG can help with collecting most relevant contexts without the need of pooling all data together in a centralized manner. Furthermore,
at inference/generation stage, C-FedRAG provides a way to utilize several specialized expert models that could offer more precise, targeted responses in niche areas. This can be achieved either by delivering precise answers by returning a specific expert's response, or combine multiple expert responses into a unified, comprehensive answer. 

Second, C-FedRAG can help multi-modality context integration. Imagine two hospitals, one with a wealth of image data, and another with extensive text data such as clinical notes and electronic medical records. Each hospital may also have its own specialized LLM fine-tuned on their respective datasets. In this scenario, C-FedRAG can merge these distinct responses via multi-modal LLMs to generate a multi-modal response. This approach combines the strengths of both datasets and the expertise embedded within each LLM, ensuring that the response is comprehensive and contextually rich, effectively capturing the nuances of each modality. 

Furthermore, C-FedRAG functions as a specialized form of a multi-site, multi-agent system, where each site in a federated environment acts as an agent capable of performing multiple tasks. These federated, multi-site, multi-task agents enable cross-country and cross-institution collaboration, harnessing collective expertise for enhanced decision-making.

\section{Conclusion}

In this paper, we introduced confidential federated RAG (C-FedRAG), a secure federated approach to Retrieval-Augmented Generation. It serves as a novel paradigm designed to securely connect and scale across decentralized networks of data providers. Conventional centralized RAG methods, while effective for reducing hallucinations and leading to more accurate retrieval, face limitations in scaling data pipelines and require centralized data repositories, which pose privacy challenges. C-FedRAG addresses these issues by leveraging federated computation with secure assurance at various stages of the RAG workflow—embedding model training, context retrieval, re-ranking, and final inference. This approach overcomes the limitations of centralized data architectures, enhancing the scalability and effectiveness of LLM applications in enterprise settings, while staying compliant with strict data access restrictions. 

Our implementation of C-FedRAG, using the NVIDIA FLARE SDK and Confidential Computing, facilitates seamless and trusted data flows between different business units and external partners. This decentralized approach improves the accuracy and relevance of LLM responses and fosters better collaboration and data sharing across organizational boundaries. The CC around the orchestrator ensures that retrieved data remains invisible to the orchestrator itself and other data providers.

As enterprises continue to explore LLM capabilities for knowledge management and information retrieval, C-FedRAG represents a fundamental shift towards decentralized and collaborative AI solutions over monolithic, data lake-oriented AI applications. By addressing the inherent challenges of data integration and access in complex enterprise environments, C-FedRAG paves the way for more effective and widespread adoption of LLM technologies, driving greater innovation and efficiency in knowledge-based tasks.

\bibliography{fedrag}
\bibliographystyle{mlsys2025}

\end{document}